\documentclass[prb,preprint,floatfix,amsmath,amssymb,superscriptaddress]{revtex4-1}
\usepackage{graphicx} 
\usepackage{bm}
\usepackage{CJK}
\usepackage{url}
\usepackage{comment}
\usepackage{lineno,hyperref}
\usepackage{subcaption}
\def\bes#1\ees{\begin{equation}\begin{split}#1\end{split}\end{equation}}
\def\be{\begin{equation}}
\def\ee{\end{equation}}
\def\bea{\begin{eqnarray}}
\def\eea{\end{eqnarray}}

\newcommand{\matrixel}[3]{\ensuremath{\left\langle #1 \vphantom{#2#3} \right| #2 \left| #3 \vphantom{#1#2} \right\rangle}}

\begin{document}

\title{Giant and negative magnetoresistances in conical magnets}
\author{Raz Rivlis}
\email{rrivlis@uwyo.edu}
\affiliation{Department of Physics and Astronomy/3905\\
1000 E. University Avenue\\ University of Wyoming\\ Laramie, WY 82071}
\author{Andrei Zadorozhnyi}
\affiliation{Department of Physics, 37th Street\\Georgetown University \\Washington, DC 20057}
\author{Yuri Dahnovsky}
\email{yurid@uwyo.edu}
\affiliation{Department of Physics and Astronomy/3905\\
1000 E. University Avenue\\ University of Wyoming\\ Laramie, WY 82071}
\date{\today}
\begin{abstract}
We study magnetotransport in conical helimagnet crystals. Spin dependent magnetoresistance exhibits dramatic properties for high and low electron concentrations at different temperatures. For spin up electrons we find negative magnetoresistance despite only considering a single carrier type. For spin down electrons we observe giant magnetoresistance due to depletion of spin down electrons with an applied magnetic field. For spin up carriers, the magnetoresistance is negative, due to the increase in charge carriers with a magnetic field. In addition, we investigate spin dependent Hall effect. If a magnetic field reaches some critical value for spin down electrons, giant Hall resistance occurs, i.e., Hall current vanishes. This effect is explained by the absence of spin down carriers.  For spin up carriers, the Hall constant dramatically decreases with field, due to the increase in spin up electron density. Because of the giant spin dependent magnetoresistance and Hall resistivity, conical helimagnets could be useful in spin switching devices.

\end{abstract}

\maketitle

\section{Introduction}

\begin{figure}
    \includegraphics[width=4truein]{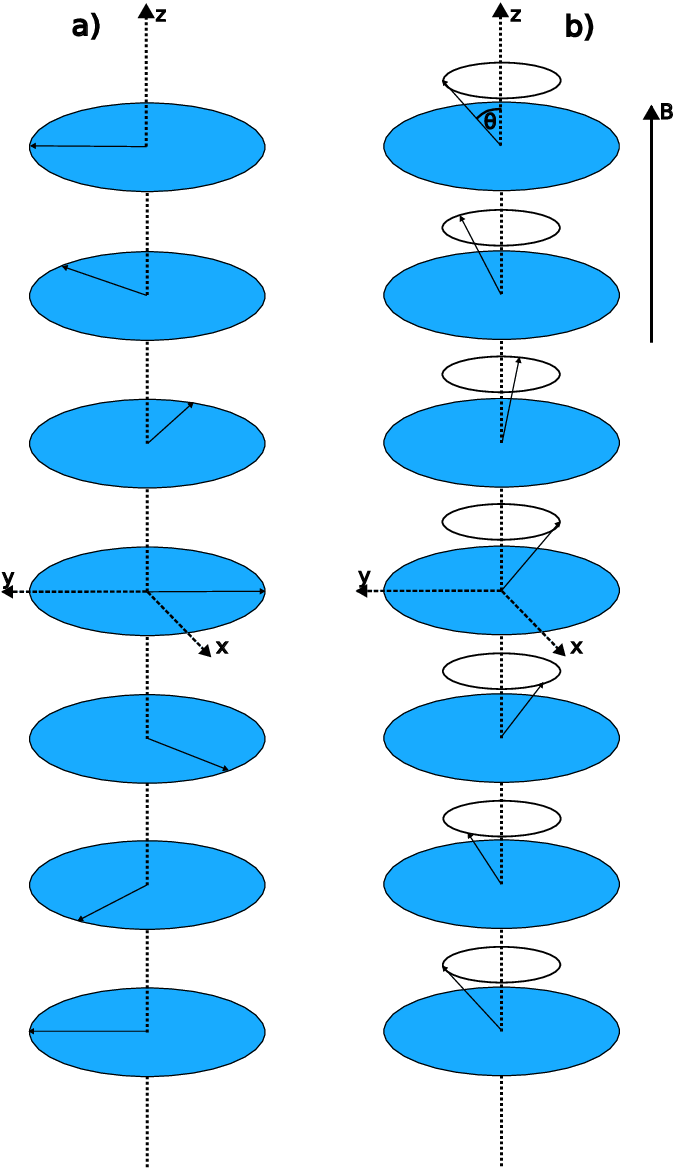}
    \caption{\small Schematic diagram of (a) helical and (b) conical magnets. Each blue circle represents a crystal plane. Each unlabeled arrow shows the alignment of the magnetic moment for each crystal plane. a) There is a helical magnetic structure where no magnetic field is applied. Magnetic moments lie in the x-y plane.  b) There is a conical magnetic structure. The magnetic field is applied in the positive z direction. The magnetic moments have a z-component.}
    \label{fig1}
\end{figure}

Helimagnetism is a phenomenon that occurs in magnetic materials and can be the result of relativistic corrections, \cite{inoue95,xiao,yazyev08, ll} frustrated magnetism\cite{sukh22}, or RKKY/Kondo effects\cite{wang20, ozawa17, hayami17, hayami21}. Conical magnetic phases exist in helimagnetic materials, along with skyrmions, ferromagnetism, etc. \cite{binz06,tsunoda89}. Conical phases were experimentally verified in $\mathrm{U_{3}P_{4}}$,\cite{sandratskii96} $\mathrm{\gamma}$-$\mathrm{Fe}$, \cite{kurz04} $\mathrm{MnP}$\cite{wang16} crystals. In the absence of a magnetic field, helimagnets can have spin spiral phases shown in Fig.\ref{fig1}a, with the spiral direction defined by the crystal anisotropy. When a magnetic field is applied and aligned with the spin spiral direction, the spins tilt towards the field direction, creating the conical phase, as shown in Fig.\ref{fig1}b. This occurs when the applied field is greater than the pinning value $B_{pin}$. \cite{nakanishi80, maleyev06, grigoriev14, nakanishi83, kulikov84, plumer90, pfleiderer05, schmidt16} In many materials $B_{pin}=0$. Spin spirals have many potential applications in spintronics. \cite{Yang21,jiang2020, mohanta20, ust, goto21} 

Our main focus in the work is transport in conical magnets. Charge transport in conical helimagnets was studied in Ref \cite{ZD4} in the absence of a magnetic field. Helical magnets in the presence of a magnetic with $B < B_{pin}$ were investigated in Ref \cite{Zadorozhnyi_2023}. In prior work we found giant magnetoresistance when the external field is applied perpendicular to the spiral propagation direction in the helical case. In addition, the Hall constant appears to be independent of the helical structure \cite{Zadorozhnyi_2023}. In this paper, we continue to study magnetotransport for $B > B_{pin}$, allowing $B_{pin} = 0$. The spin component along the spiral axis is proportional to the applied magnetic field, i.e., $S_z \propto B$. To study transport properties in these materials, we employ the kinetic Boltzmann equation with electron-acoustic phonon scattering, and then, calculate spin dependent magnetoresistance and spin dependent Hall effect.

Transport properties in conical magnetic materials can be described by the Hamiltonian \cite{kurz04}
\begin{align}
\label{h0}
\hat{H}_0 &= \hat{H}_{crys} + \hat{H}_{hel} = \frac{\hbar^2 k^2}{2m} - J\bm{S}\cdot \hat{\bm{\sigma}} \\
&= \frac{\hbar^2 k^2}{2m} - JS_{\|}\left(\sigma_{x} \cos(\varkappa z) + \sigma_{y} \sin(\varkappa z)\right) - JS_z\sigma_{z},
\end{align}
where $\bm{\sigma}$ is the vector of the three Pauli matrices, $J$ is the the interaction energy, $\varkappa$ is the spin spiral wavenumber, $m$ is the electron effective mass, and $\bm{S}$ is the magnetic moment of the crystal. $S_z$ is completely induced by the magnetic field and defined as
\begin{align}
\label{sz}
S_z &= \beta B,
\end{align}
where $\beta$ is a constant dependent on the material. $S_\|$ is the competent of $S$ in the xy plane.
This Boltzmann equation approach was applied to charge transport in helical, conical, and skyrmion phases. \cite{kurz04, sandratskii86, kurebayashi21,bruno04, sitte14}

\section{Theory}
\subsection*{Electronic Structure}

Hamiltonian (\ref{h0}) was exactly diagonalized in Ref \cite{ZD4} where the two energy bands are

\bes \label{eps}
 \varepsilon_{1, 2} &=  \frac{\varepsilon_{0} \left(\bm{k} + \bm{e}_{z} \frac{\varkappa}{2} \right) + \varepsilon_{0} \left(\bm{k} - \bm{e}_{z} \frac{\varkappa}{2} \right)}{2} 
\pm \sqrt{ J ^ {2} S _{\parallel} ^ {2} + \left[ \frac{\varepsilon_{0} \left(\bm{k} + \bm{e}_{z} \frac{\varkappa}{2} \right) - \varepsilon_{0} \left(\bm{k} - \bm{e}_{z} \frac{\varkappa}{2} \right)}{2} + J S_{z}\right] ^ { 2 } } \\
& = \frac{\varepsilon_{0} \left(\bm{k} + \bm{e}_{z} \frac{\varkappa}{2} \right) + \varepsilon_{0} \left(\bm{k} - \bm{e}_{z} \frac{\varkappa}{2} \right)}{2} 
\pm \sqrt{ J ^ {2} S _{\parallel} ^ {2} + \left[ D + J S_{z}\right] ^ { 2 } },
\ees
where $D = \left[\varepsilon_{0} \left(\bm{k} + \bm{e}_{z} \frac{\varkappa}{2} \right) - \varepsilon_{0} \left(\bm{k} - \bm{e}_{z} \frac{\varkappa}{2} \right)\right]/2$

The wavefunctions can be written as follows:
\bes \label{ef}
\left(
\begin{matrix}
	\Psi_{\bm{k}, \varkappa} ^{\uparrow \left(\nu\right) } \left( \bm{r} \right)\\
	\Psi_{\bm{k}, \varkappa} ^{\downarrow \left(\nu\right) } \left( \bm{r} \right)
\end{matrix}
\right) 
= 
\left(
\begin{matrix}
	a _ { \nu } \left( \bm{k}, \varkappa \right) e^{ - i \frac{\varkappa}{2} z } \\
	b _ { \nu } \left( \bm{k}, \varkappa \right) e^{ + i \frac{\varkappa}{2} z }
\end{matrix}
\right)
 \psi_{0 \bm{k}} \left( \bm{r} \right),
\ees
where $\nu$ represents the band index and

\bes \label{ab2}
a_{1} & = b_{2}  = \frac{1}{\sqrt{2}}\frac{ \sqrt{ \sqrt{ J ^ {2} S _{\parallel} ^ {2} + \left(  D + J S_{z}\right) ^ { 2 } } +  \left(  D + J S_{z}\right) } }{ \left( J ^ {2} S _{\parallel} ^ {2} + \left(  D + J S_{z}\right) ^ { 2 } \right)^{1/4} },\\
a_{2} & = - b_{1} = - \frac{1}{\sqrt{2}}\frac{ \sqrt{ \sqrt{ J ^ {2} S _{\parallel} ^ {2} + \left(  D + J S_{z}\right) ^ { 2 } }  -  \left(  D + J S_{z}\right) } }{ \left( J ^ {2} S _{\parallel} ^ {2} + \left(  D + J S_{z}\right) ^ { 2 } \right)^{1/4} }.
\ees
\begin{figure}[!ht]
    \includegraphics[width=\linewidth]{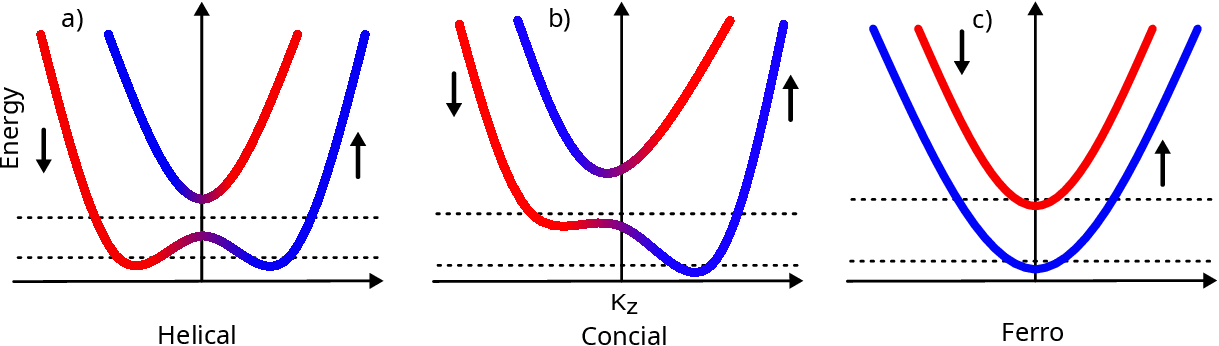}
    \caption{\small Relation between a band structure and a magnet type. The blue and red lines denote spin up and spin down electrons, respectively. The two dashed lines correspond to  chemical potentials for low and high electron concentrations.  We schematically depict  the band structure for (a) helical materials ($B=0$), (b) conical magnets, and (c)  ferromagnets, respectively ($B$ is strong).}   
\label{fig2}
\end{figure}

According to Eq. (\ref{sz}), the energy bands defined by Eq. (\ref{eps}), and wavefunctions defined by Eqs. (\ref{ef}) and (\ref{ab2}), all depend on magnetic field. As shown below, this field dependence will dramatically change magnetoresistance and Hall effect. The band structure is depicted in Fig. \ref{fig2}. Red and blue lines correspond spin up and spin down electrons, respectively. Fig. \ref{fig2}a represents a helical band structure with no magnetic field ($S_z = 0$, $S_\| \neq 0$). Fig. \ref{fig2}b shows the conical case ($S_z \neq 0$, $S_\| \neq 0$). Fig. \ref{fig2}c shows the ferromagnetic case at larger magnetic fields ($S_z \neq 0$, $S_\| = 0$). The dashed lines approximately correspond to low and high Fermi levels where the electron concentration is constant and independent of magnetic field.

\subsection*{Transport}

For transport calculations, we consider the Boltzmann equation with relaxation rate due to electron-phonon interaction: \cite{anselm}
\begin{equation}
\label{bol}
- \frac{e}{\hbar} \left( \bm{E} + \left[ \bm{v}^{\nu} \times \bm{B} \right] \right) \bm{\nabla}_{\bm{k}} \left( f_{0}\left(\varepsilon _{\nu}\left(\bm{k}\right) \right) + f_{1}^{\nu}(\bm{k}) \right)=\sum_{\nu ^{\prime}}\sum_{\bm{k}^{\prime}}\left(W_{\bm{k}\bm{k}^{\prime}}^{\nu \nu ^{\prime}}f_{1}^{\nu ^{\prime}}(\bm{k}^{\prime})-W_{\bm{k}^{\prime}\bm{k}}^{\nu ^{\prime} \nu}f_{1}^{\nu}(\bm{k})\right).
 \end{equation}
$f_0(k)$ is the equilibrium Fermi distribution function, $f_{1}(k)$ is the nonequilibrium part of the total distribution function, $\bm{E}$ is an applied electric field, $\bm{B}$ is an applied magnetic field, $\bm{v}$ is an electron velocity, and $ \bm{\nabla}_{\bm{k}} $ is the gradient with respect to the wavevector. 
 
The transition rates $W_{\bm{k}\bm{k}^{\prime}}^{\nu \nu ^{\prime}}$ are defined as follows:
\be \label{w0}
W_{\bm{k} \bm{k}^{\prime}}^{\nu\nu^\prime}=(2\pi/\hbar)\left|\matrixel{\bm{k}^{\prime}, \nu^{\prime}, N_{\bm{q}j}^{\prime}}{\Delta V}{\bm{k}, \nu, N_{\bm{q}j}}\right|^{2} \delta(\varepsilon_{\nu}(\bm{k}) - \varepsilon_{\nu^{\prime}}(\bm{k}^{\prime})).
\ee
where $\Delta V$ is the electron-acoustic phonon interaction potential. $N_{\bm{q}j}$ is the population number of phonons with the wavevector $\bm{q}$ and the branch $j$ determined from the Bose distribution function:
\be\label{bose}
N_{\bm{q}j} = \frac{1}{e^{\frac{\varepsilon_{ph}}{k_{B}T}} - 1}.
\ee
Index $\nu $ denotes an energy band number. 

For pure crystals, the electron scattering is only from electron-phonon interactions. In our calculations we only consider acoustic phonons. Then the electron-phonon interaction yields:
  \be \label{eph}
 \hat{V}_{e-ph} \approx - \bm{\nabla} \left(\hat{H}_{crys} + \hat{H}_{hel}\right) \cdot \bm{u}.
 \ee
 $\hat{H}_{crys}$ and $\hat{H}_{hel}$ are defined in Eq. (\ref{h0}). The atom displacement, $\bm{u}$,  can be expressed in terms of normal phonon coordinates. The transition rates in Eq. (\ref{w0}) are determined for electron and phonon wavefunctions separately. Note, since $\hat{H}_{hel}$ in Eq.(\ref{eph}) is spin dependent, the transition rates are also spin dependent.
 
Electric current density can be found from the nonequilibrium distribution function:
\be\label{j}
j_{i}^{\nu} = e \frac{1}{ (2 \pi) ^{3} } \int f_{1}^{\nu} v_{i}^{\nu} d^{3} \bm{k} ,
\ee
where $v_{i}^{\nu}$ is a velocity projection ($i = x, y, z$) determined as $v_{i}^{\nu} = \partial \varepsilon^{\nu}(\bm{k}) / \hbar \partial k_{i} $. It is important to note that the magnetic field dependence comes into current density through the nonequilibrium distribution function and the electron velocities. In particular, the distribution function depends on the kinetic part of the Boltzmann equation (\ref{bol}), the transition rates (\ref{w0}), the energy bands (\ref{eps}), and wavefunctions (\ref{ef}).

\subsection*{Computational Details}

The Boltzmann Eq. (\ref{bol}) is solved numerically using original code. The relaxation rates are considered within the first Born approximation with respect to electron-acoustic phonon interaction. Eq. (\ref{bol}) is a linear integro-differential matrix equation for $f_{1}$. For calculations, it is important  remove the derivative, $\bm{\nabla}_{\bm{k}} f^{\nu}_{1}$. The procedure is described in detail in Ref.\cite{Zadorozhnyi_2023} Keeping electron concentration constant, we calculate the chemical potential $\mu$, which is $B$ dependent because of the magnetic field dependence of the energy bands as shown in Eqs. (\ref{eps}).

For the calculations we choose the following values: $m = 0.5$, $J = 0.02$, $|S| = 1$, small $n = 0.002275$ at $T = 10K$, large $n = 0.01407$ at $T = 10K$, small $n = 0.002780$ at $T = 90K$, and large $n =0.01477$ at $T = 90K$. $\beta$ is chosen such that the conical regime is between 0 and 10 Tesla. The spin spiral period is 7 lattice lengths of $L = 0.526$nm, corresponding to bcc $Eu$ metal  \cite{olsen64}.

\section{Results}

In this section we present the calculated spin dependent magnetoresistance and Hall current, depending on magnetic field, electron concentration, and temperature. 

\subsection*{Spin dependent magnetoresistance}

\begin{figure}
    \centering
    \begin{subfigure}{.45\textwidth}
    	\centering
    	\includegraphics[width=\linewidth]{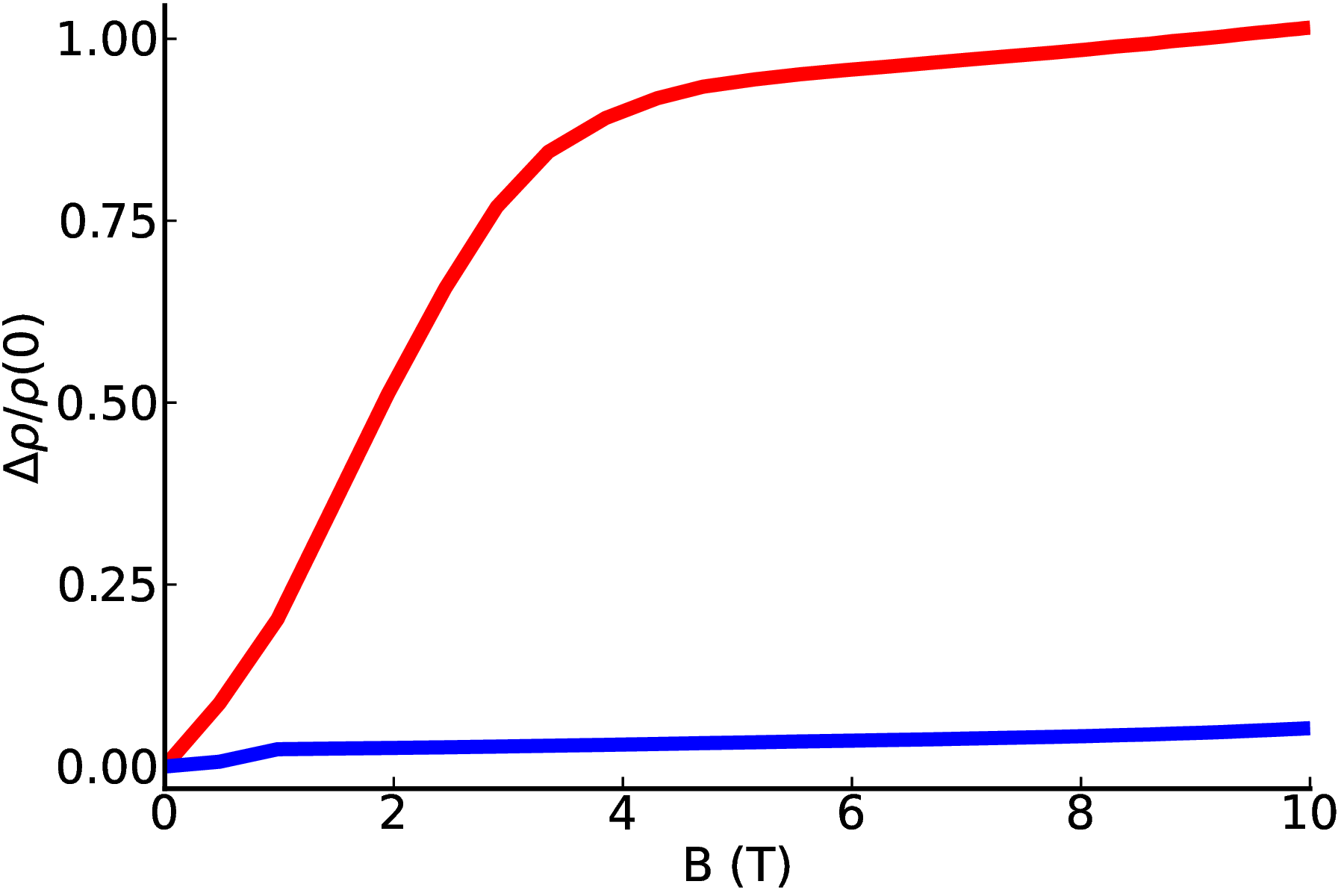}
    	\caption{\small Total magnetoresistance at 10K}
    	\label{mra}
    \end{subfigure}
    \begin{subfigure}{.45\textwidth}
    	\centering
    	\includegraphics[width=\linewidth]{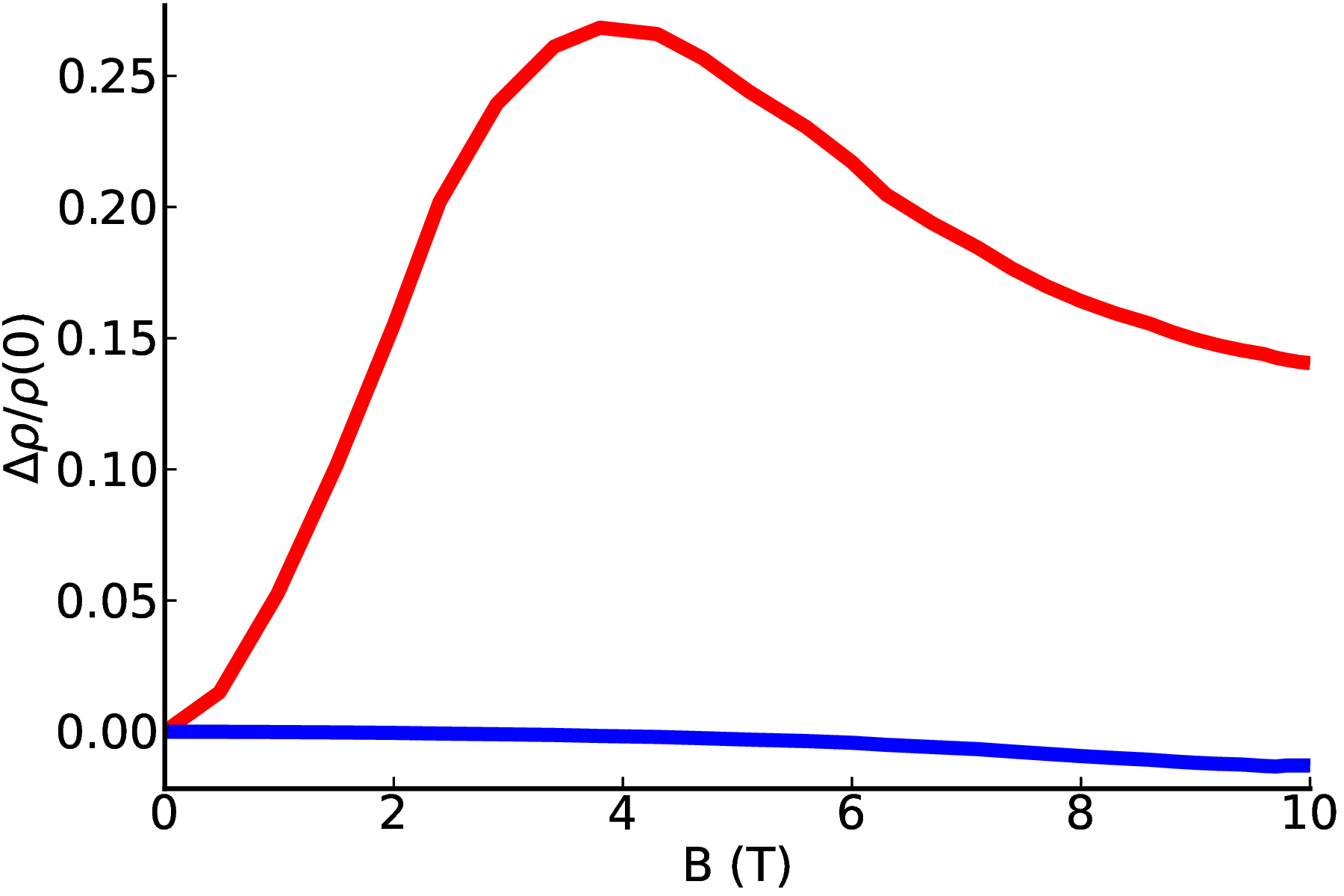}
    	\caption{\small Total magnetoresistance at 90K}
    	\label{mrb}
    \end{subfigure}
    \caption{\small Total magnetoresistance in a conical magnet. The red and blue lines represent magnetoresistance at low and  electron concentrations, respectively, at (a) $T= 10$K, and (b) $T= 90$K} \label{mr}
\end{figure}

In Fig. \ref{mr} we present the total (spin up and spin down) magnetoresistance ($\Delta \rho / \rho(0)$) as a function of magnetic field. We show the dependence of $\Delta \rho / \rho(0)$ on magnetic field at two different electron concentrations, at $T =10K$, Fig. \ref{mra}, and at $T = 90K$, Fig. \ref{mrb}. The blue line defines large electron concentration ($e_F >> k_B T$). The red line corresponds to small $n$ ($\varepsilon_F < k_B T$). According to Askerov \cite{askerov70}, at $T = 0$, $\Delta \rho / \rho \rightarrow 0$ for all magnetic fields. Because $\varepsilon_F >> k_B T$ (the low temperature limit) the blue lines at both temperatures, Fig. \ref{mra} and Fig. \ref{mrb}, are close to zero. For lower n's ($\varepsilon_F < k_B T$), we have a high temperature limit, and at small field, according to Anselm \cite{anselm}, $\Delta \rho / \rho \propto H^2$. At large values of $B$, the magnetoresistance is field independent. The unexpected local maximum in Fig. \ref{mrb} is due to spin effects described below.

\begin{figure}
    \centering
    \begin{subfigure}{.45\textwidth}
    	\centering
    	\includegraphics[width=\linewidth]{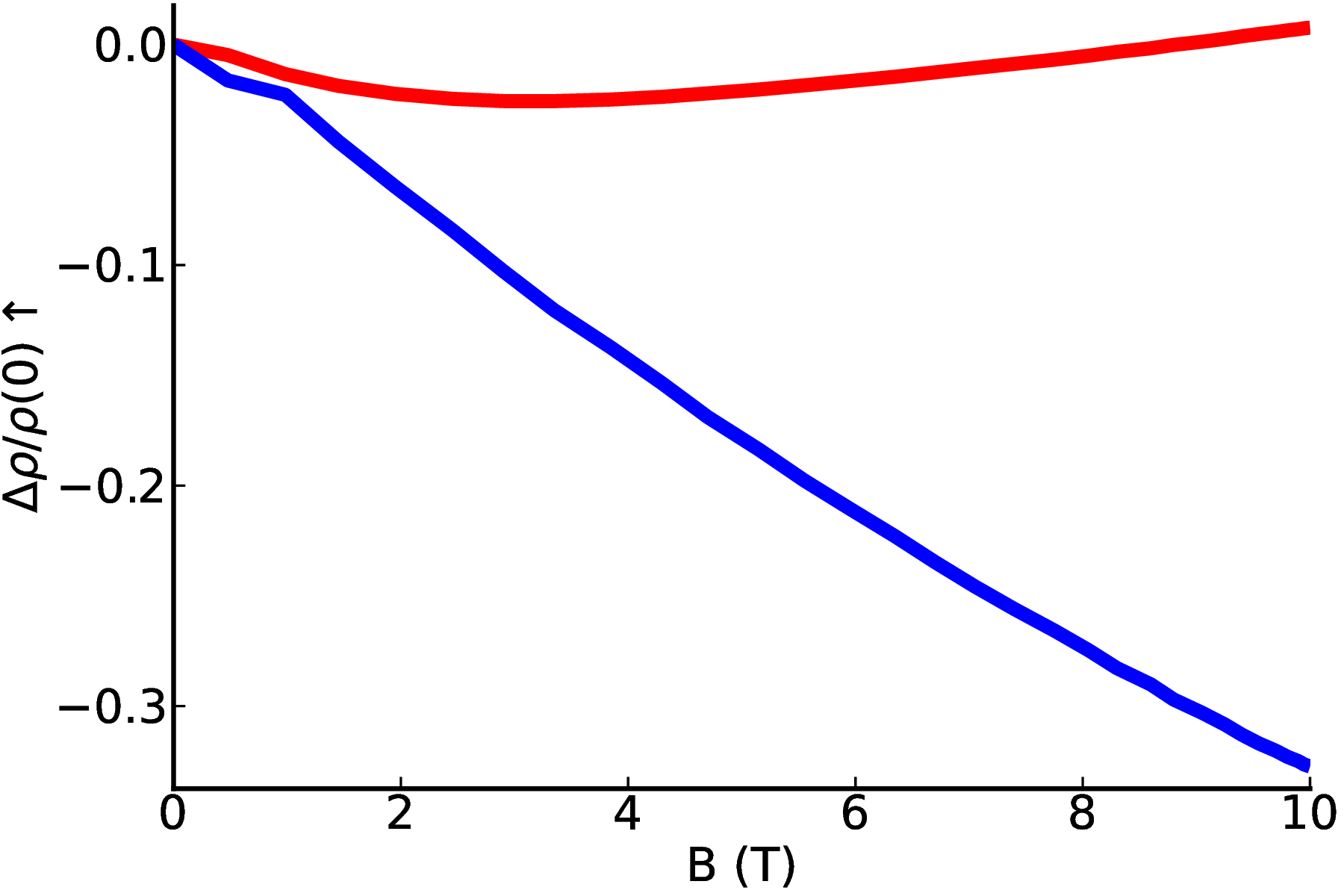}
    	\caption{\small Spin up magnetoresistance at 10K}
    	\label{spinupa}
    \end{subfigure}
    \begin{subfigure}{.45\textwidth}
    	\centering
    	\includegraphics[width=\linewidth]{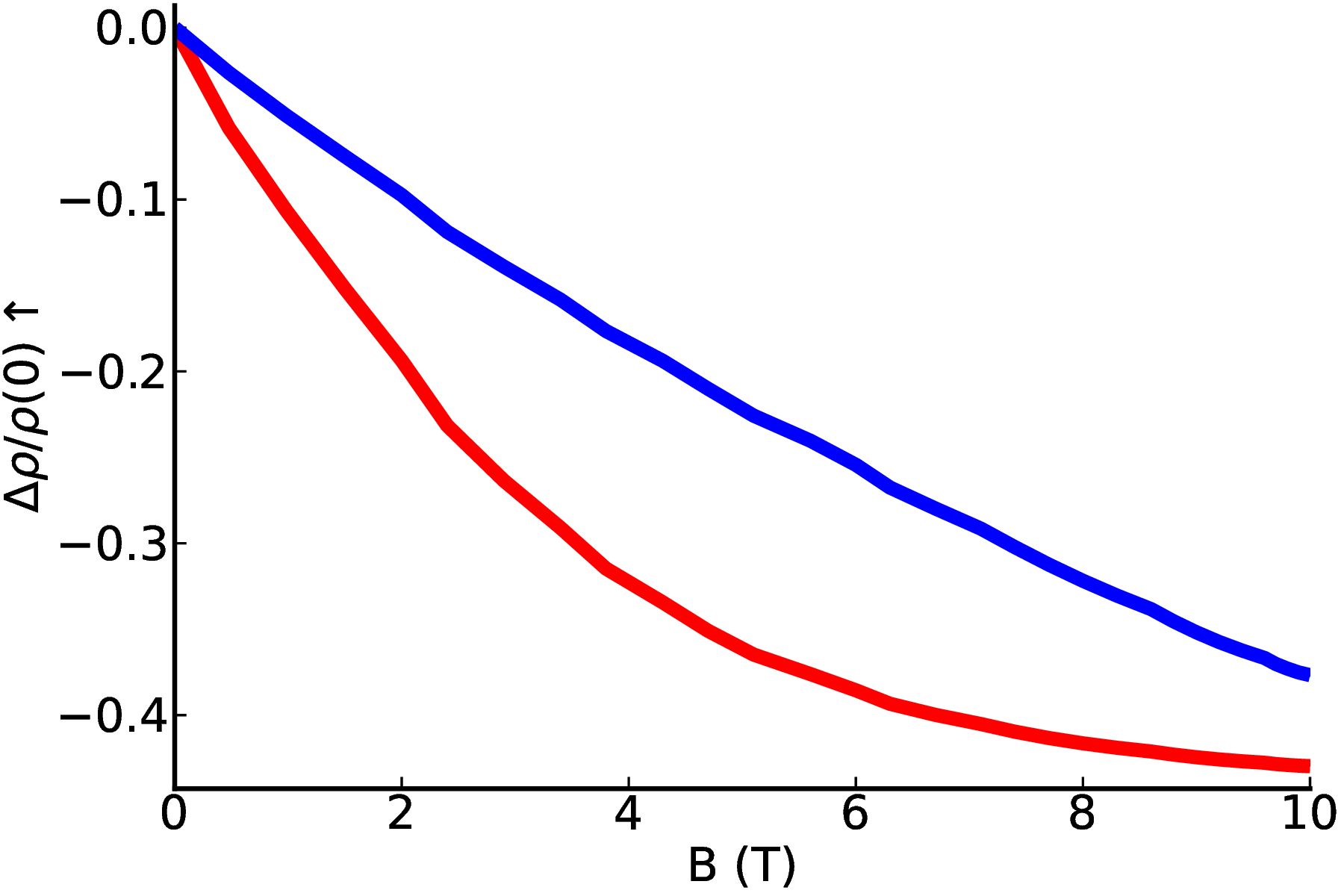}
    	\caption{\small Spin up magnetoresistance at 90K}
    	\label{spinupb}
    \end{subfigure}
    \caption{\small Spin up magnetoresistance in a conical magnet. The red  and blue lines represent magnetoresistances at low and high electron concentrations, respectively at temperatures (a) 10K, and (b) 90K}
    \label{spinup}
\end{figure}

\begin{figure}    
    \centering
    \begin{subfigure}{.45\textwidth}
    	\centering
    	\includegraphics[width=\linewidth]{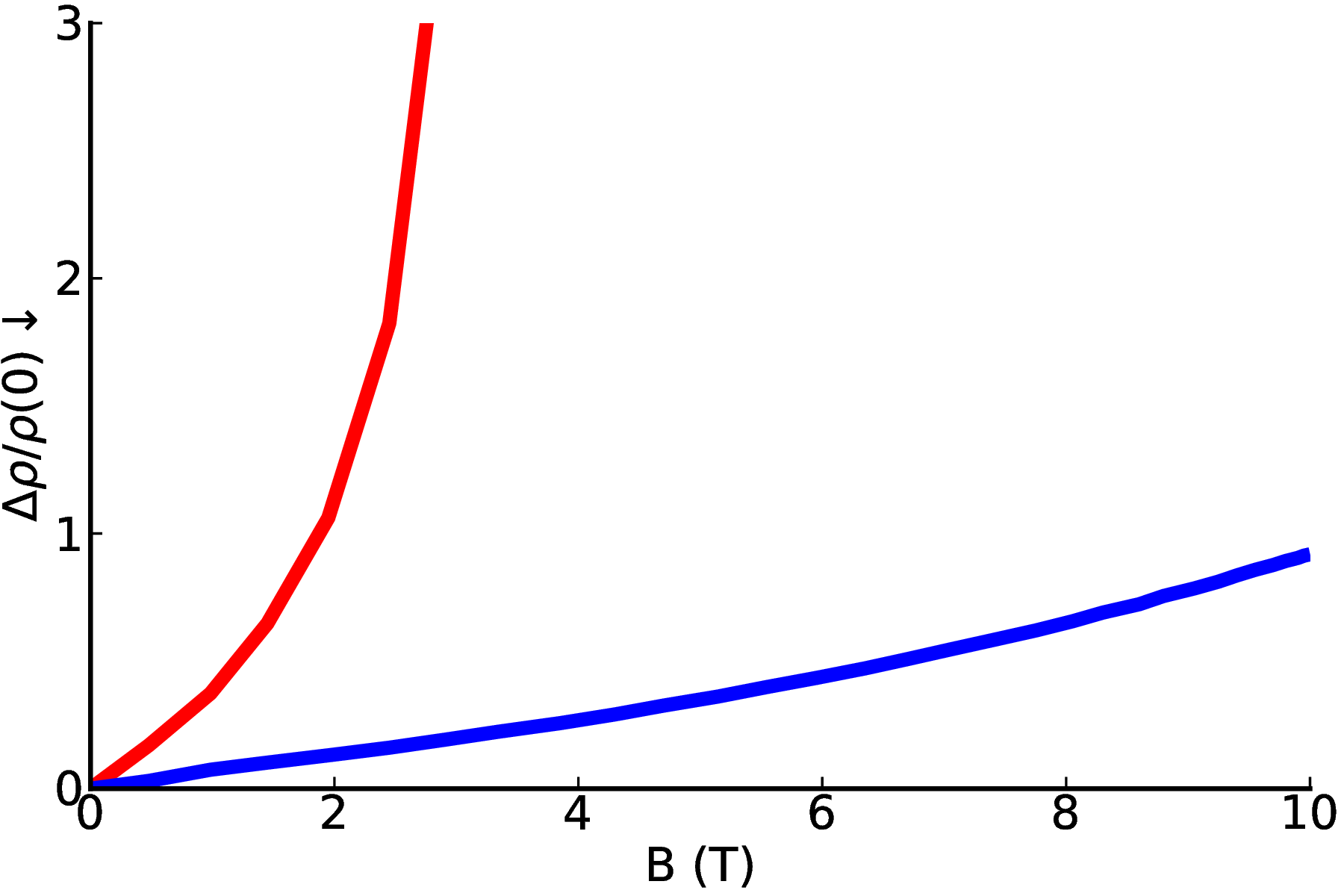}
    	\caption{\small Spin down magnetoresistance at $T=10$K}
    	\label{mrdowna}
    \end{subfigure}
    \begin{subfigure}{.45\textwidth}
    	\centering
    	\includegraphics[width=\linewidth]{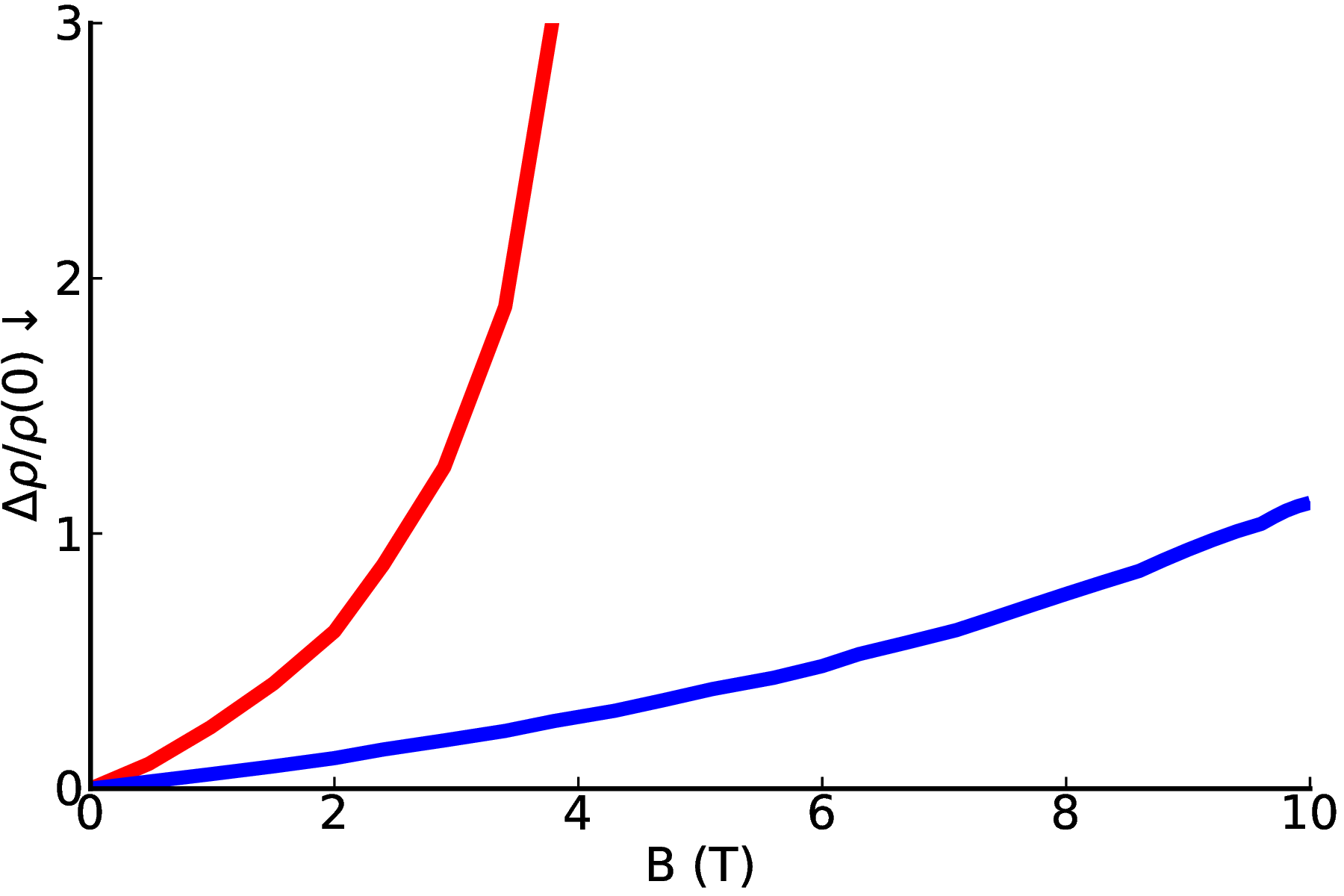}
    	\caption{\small Spin down magnetoresistance at $T=90$K}
    	\label{mrdownb}
    \end{subfigure}
    \caption{\small Spin down magnetoresistance in a conical magnet. The red  and blue lines represent magnetoresistances at low and high electron concentrations, respectively at temperatures (a) 10K, and (b) 90K}
    \label{spindown}
\end{figure}

The magnetoresistance as a function of magnetic fields exhibits dramatic behavior when we consider spin up and spin down contributions separately as shown in Fig. \ref{spinup} and Fig. \ref{spindown}. Typically, negative magnetoresistance can occur when the primary carrier changes charge sign, i.e., from electrons to holes or vice versa. Our only carriers are electrons, so one would expect $\Delta \rho / \rho(0) > 0$. However, as shown in Fig. \ref{spinup}, we find that magnetoresistance is negative in both high and low temperature limits. This happens because the band structure shown in Fig. \ref{fig2}, changes with $B$ according to Eq. (\ref{eps}). Because $\Delta \rho = \rho(B) - \rho(0)$, negative magnetoresistance occurs when $\rho(B) < \rho(0)$. Due to the energy band well shift, the spin up electron concentration, $n_\uparrow(B)$, increases, while the spin down concentration, $n_\downarrow(B)$ decreases. Conductivity increases with electron concentration, therefor decreasing resistivity ($\rho(B)_\uparrow$, $\rho \sim \sigma^{-1}$). The total electron concentration, $n(0) = n(B) = n_\uparrow(B) + n_\downarrow(B)$, is conserved. Thus, $\rho(B)_\uparrow < \rho(0)_\uparrow$ and $\Delta \rho_\uparrow$ is negative. The temperature dependence shown in Fig. \ref{spinupa} and \ref{spinupb} can be explained as follows: for high electron concentrations, the blue lines, are independent of temperature, $T = 10K$ and $T = 90K$. According to Ref \cite{askerov70}, $\Delta \rho / \rho(0)$ vanishes in the low temperature limit. However, we do not observe this dependence due to the spin contributions described above. For low electron concentrations, the red lines, $|\Delta \rho_\uparrow / \rho(0)_\uparrow |_{T=10} << |\Delta \rho_\uparrow / \rho(0)_\uparrow |_{T=90}$ in the high temperature limit according to Ref \cite{askerov70}.

We now demonstrate the magnetoresistance dependence on $B$ for the spin down electrons in Fig. \ref{spindown}. In both Figs. \ref{mrdowna} and b, for low total electron concentration, the red lines exhibit giant magnetoresistance. This giant magnetoresistance can be explained by decreasing spin down electron concentration. Such a change occurs because of the change in band structure that goes to a ferromagnetic state as shown in Fig. \ref{fig2}. In this case, the Fermi level is near the bottom of the lower band (spin up band), therefor, $n_\downarrow (B) \rightarrow 0$, and electroconductivity decreases, $\rho(B)_\downarrow >> 1$. Consequently, $(\rho(B)_\downarrow - \rho(0)_\downarrow) / \rho(0)_\downarrow >> 1$. For large electron concentrations and magnetic fields (blue lines) the energy structure turns into the ferromagnetic state shown in Fig. \ref{fig2}c. In this case, $\varepsilon_F$ is slightly above the minimum of the upper band, where the spin down electron concentration is low, but nonzero, causing the increase in resistivity.

\subsection*{Spin dependent Hall constant}

\begin{figure}
    \centering
    \begin{subfigure}{.45\textwidth}
    	\centering
    	\includegraphics[width=\linewidth]{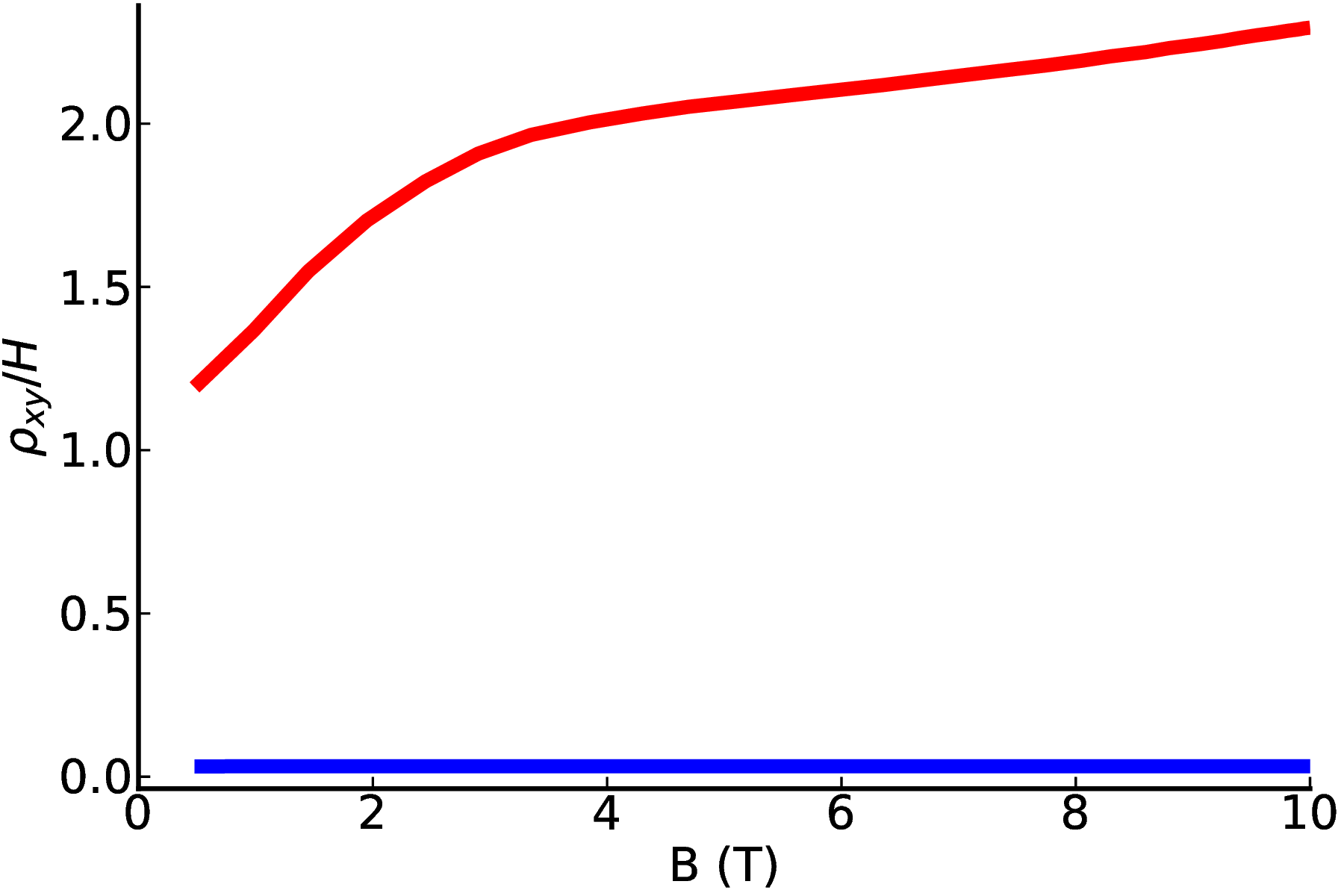}
    	\caption{\small Total Hall constant at $T=10$K}
    	\label{Halla}
    \end{subfigure}
    \begin{subfigure}{.45\textwidth}
    	\centering
    	\includegraphics[width=\linewidth]{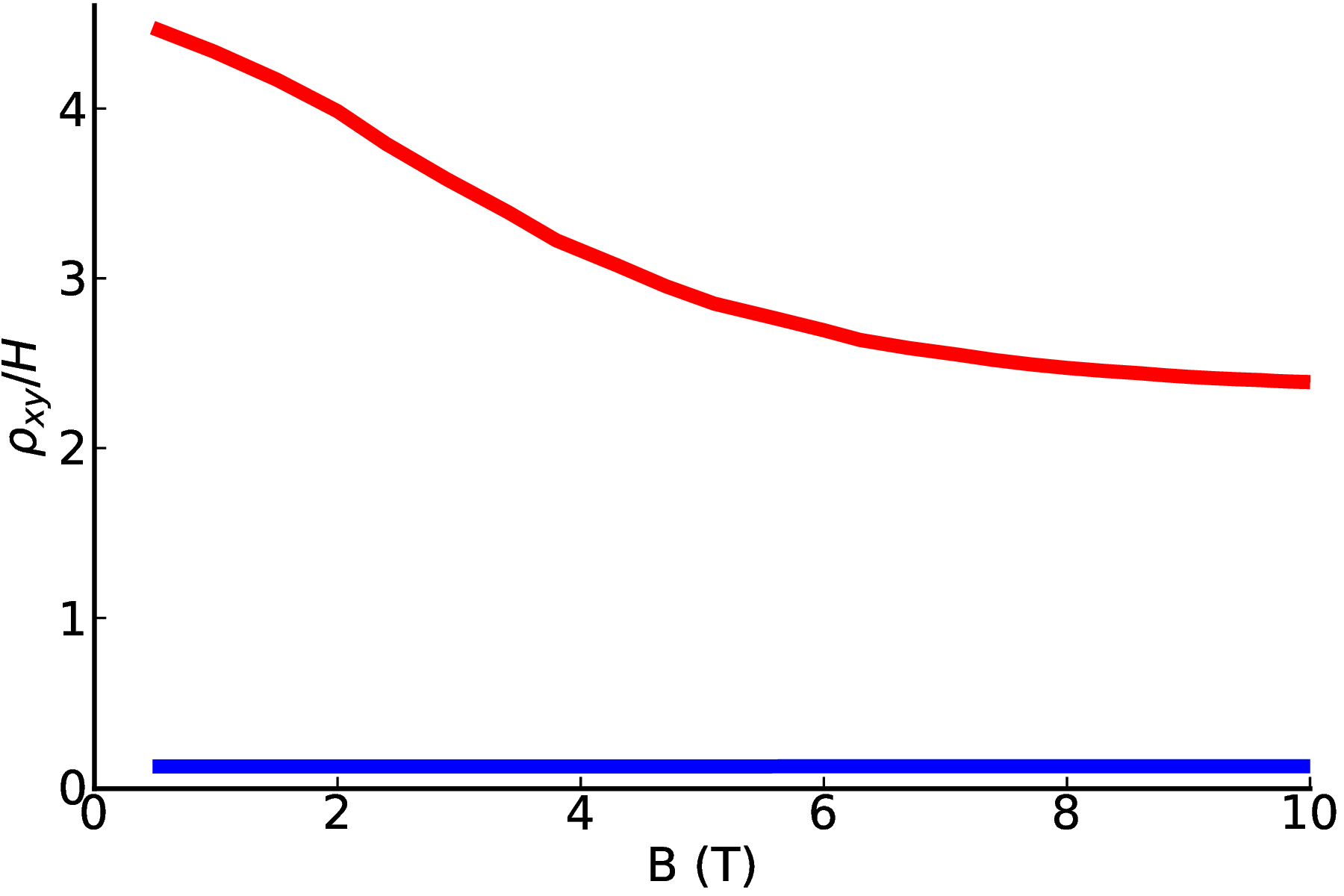}
    	\caption{\small Total Hall constant at $T=90$Kt}
    	\label{Hallb}
    \end{subfigure}
    \caption{\small Total Hall constant in a conical magnet. Red and blue lines represent Hall constant at low and high electron concentrations, respectively at (a) $T=10$K  and (b) $T=90$K.}
    \label{hall}
\end{figure}

Besides magnetoresistance, we also study spin dependent Hall effect defined in the following way:\cite{anselm}

\be\label{heff}
\rho_{xy}=-RB.
\ee
For free electrons Hall constant becomes
\be\label{Rfree}
R=\frac{1}{cen}.
\ee
As shown in Ref. \cite{Zadorozhnyi_2023}, helicity does not change the free electron Hall constant in a helical phase as described by Eq. (\ref{Rfree}). In a conical phase, the situation is different because the band structure depends on magnetic field. Thus, we expect more complicated dependences as shown in Fig. \ref{hall}. Indeed, the total Hall constant becomes field dependent, contrary to the free electron case (see Eq. (\ref{Rfree})). Fig. \ref{hall} demostrates the total Hall constant at low and high electron concenrations at $T = 10K$ and $T = 90K$. For high electron concentrations at both temperatures (blue lines), the Hall constant is small, in accordance with Eq. (\ref{Rfree}). According to Eq. (\ref{Rfree}), $R$ is field independent. However, in Fig. \ref{hall} we find that R depends on magnetic field for small electron concentrations. It happens because of complicated interference between spin effects and the field dependent band structure.

\begin{figure}
    \centering
    \begin{subfigure}{.45\textwidth}
    	\centering
    	\includegraphics[width=\linewidth]{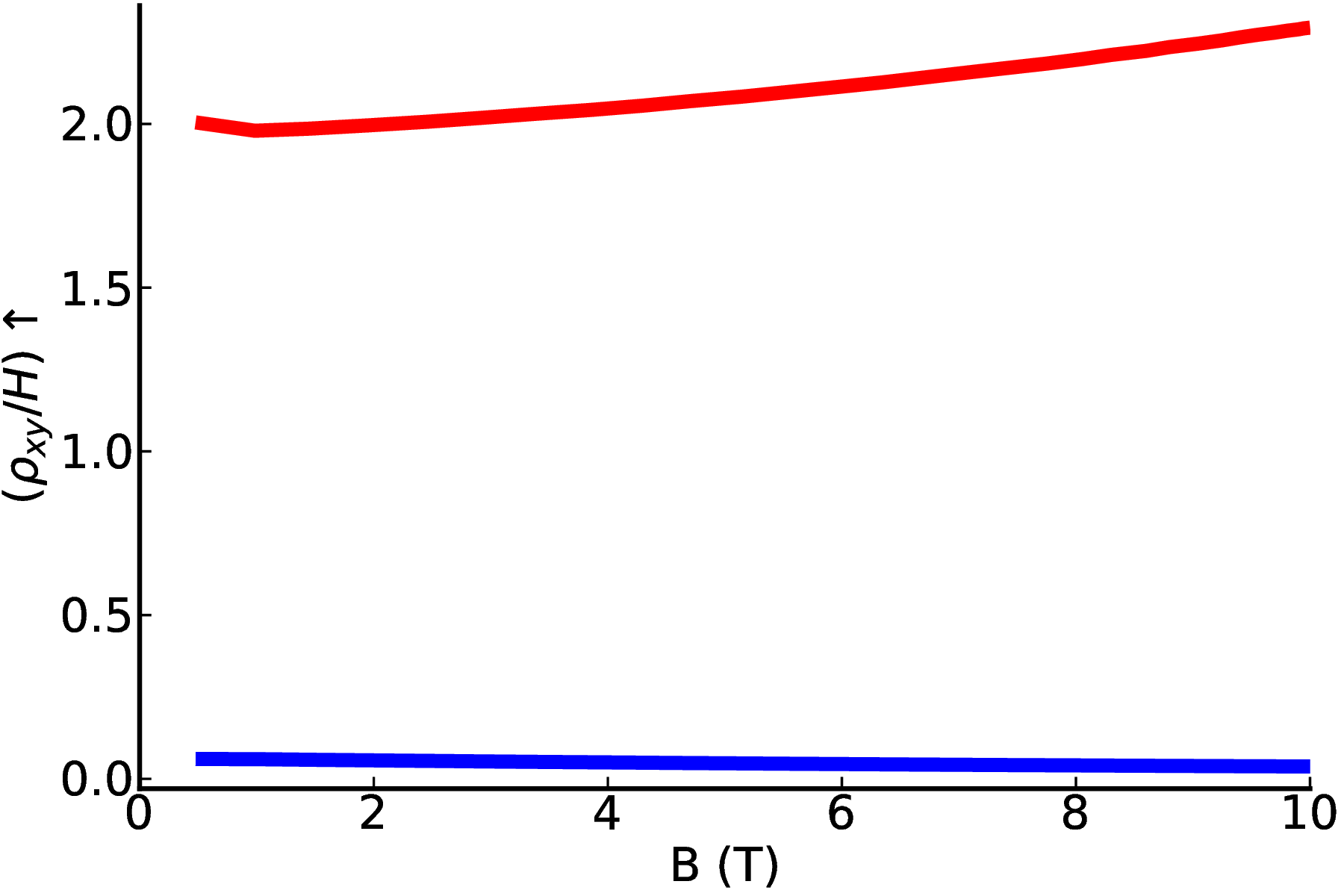}
    	\caption{\small Spin up Hall constant at $T=10$K}
    	\label{spinhalla}
    \end{subfigure}
    \begin{subfigure}{.45\textwidth}
    	\centering
    	\includegraphics[width=\linewidth]{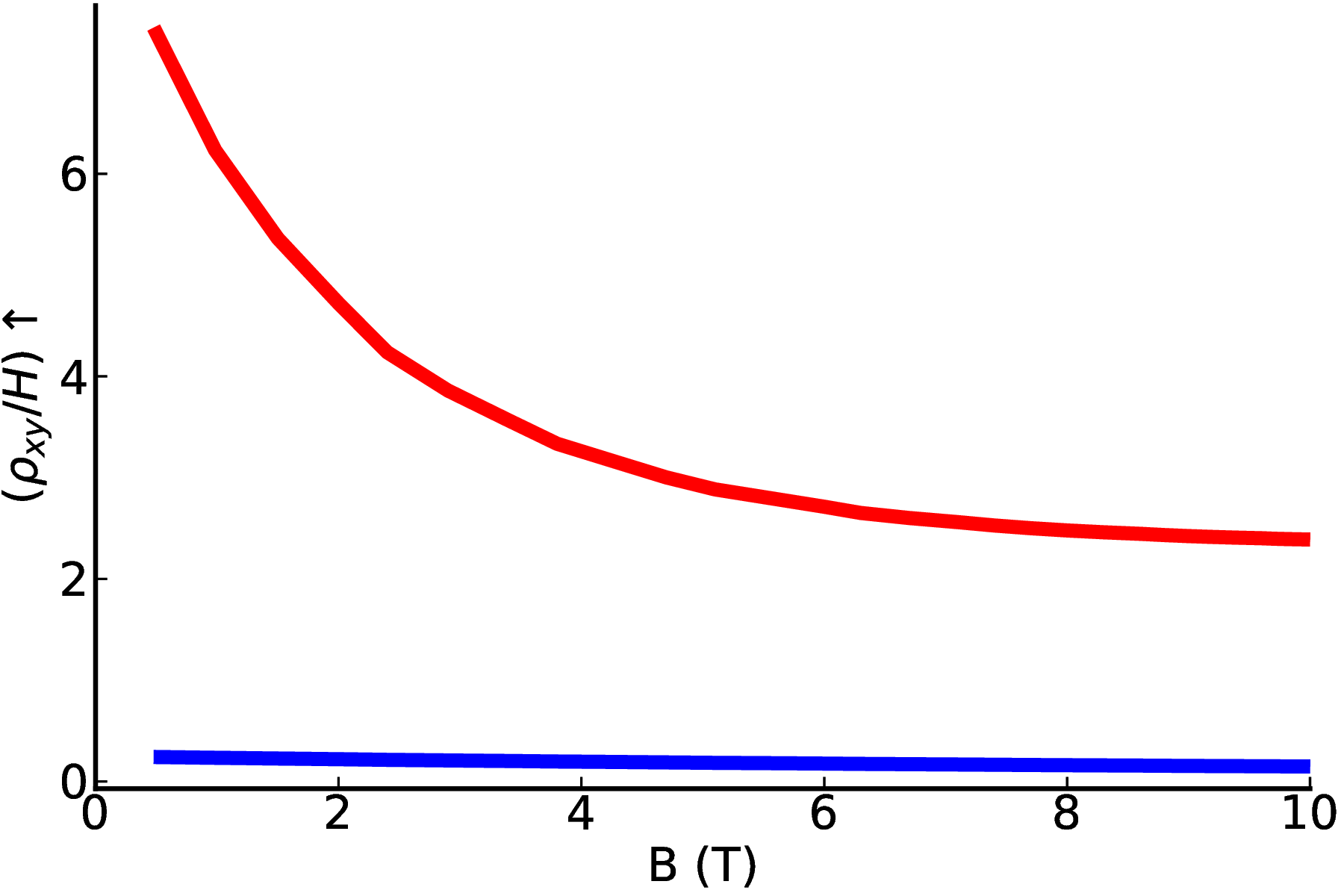}
    	\caption{\small Spin up Hall constant at $T=90$K}
    	\label{spinhallb}
    \end{subfigure}
    \caption{\small Spin up Hall constant in a conical magnet. Red and blue lines represent the Hall constant at low and high electron concentrations, respectively at (a) $T=10$K  and (b) $T=90$K}
    \label{spinuphall}
\end{figure}

\begin{figure}
    \begin{subfigure}{.45\textwidth}
    	\centering
    	\includegraphics[width=\linewidth]{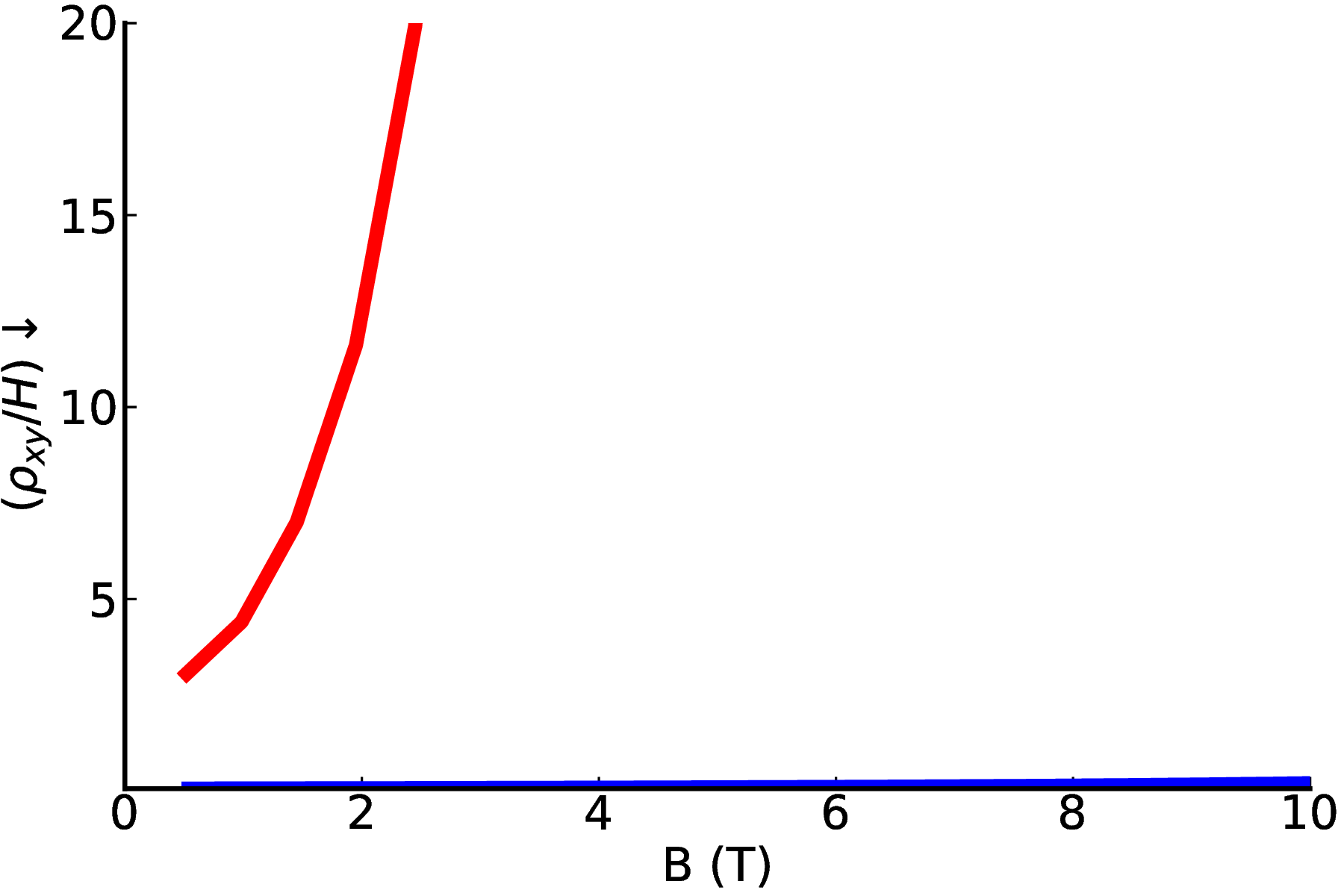}
    	\caption{\small Spin down Hall constant at $T=10$K}
    	\label{spinhallc}
    \end{subfigure}
    \begin{subfigure}{.45\textwidth}
    	\centering
    	\includegraphics[width=\linewidth]{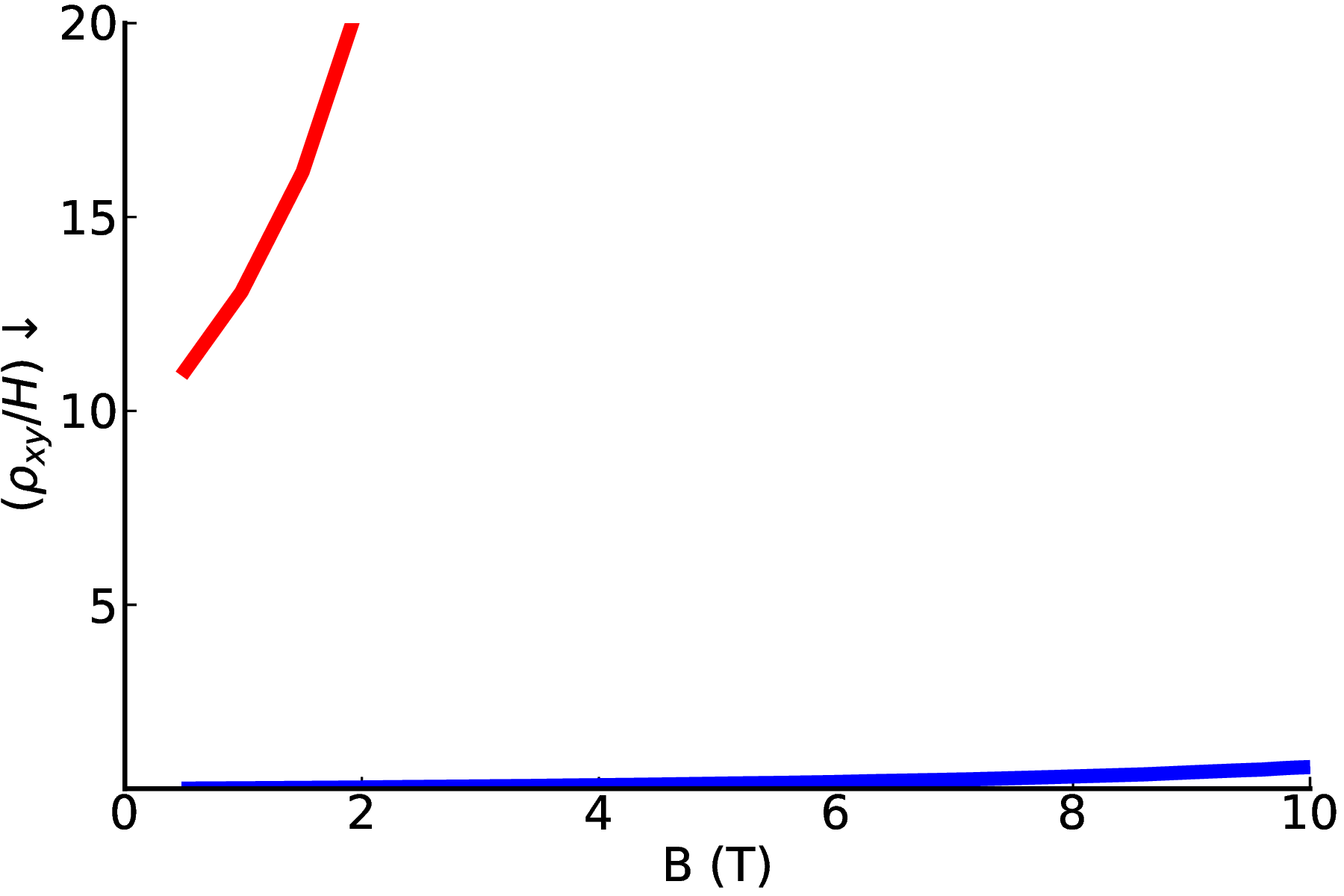}
    	\caption{\small Spin down Hall constant at $T=90$K}
    	\label{spinhalld}
    \end{subfigure}
    \caption{\small Spin down Hall constant in a conical magnet. Red and blue lines represent the Hall constant at low and high electron concentrations, respectively at (a) $T=10$K  and (b) $T=90$K}
    \label{spindownhall}
\end{figure}

To provide deeper insight, we study Hall resistivity for spin up and spin down electrons separately as shown in Figs. \ref{spinuphall} and \ref{spindownhall}. In Fig. \ref{spinuphall}, we present the dependence of the Hall constant for spin up electrons at low and high electron concentrations, and at $T = 10K$ and $T = 90K$. In Fig. \ref{spinhallb} ,the Hall constant decreases with magnetic field. This effect can be explained by the increase in spin up electron concentration ($n_\uparrow$) with magnetic field. However, for low temperature, $R$ is almost independent of $B$ as shown in Fig. \ref{spinhalla}. This effect can be explained by the competition between spin up electron concentration and the band structure change. For large electron concentrations (blue lines), $R$ is much smaller than that of the small electron concentrations, as described by Eq. (\ref{Rfree}).

In Fig. \ref{spindownhall}, we present the Hall constant for spin down electrons with respect to magnetic field. For both $T = 10K$ and $T = 90K$ (Fig. \ref{spinhallc} and Fig. \ref{spinhalld} respectively), and small electron concentrations, we find giant Hall constant, indicating the absence of spin down Hall current. However, for lower fields, the spin down Hall effect is still occurs. Such behavior can be explained by spin down electron concentration decreasing to zero. It happens because of the changing electron band as shown in Fig \ref{fig2}. In the case of large electron concentrations, the Hall constant exhibits much less dramatic behavior. The spin down electron concentration decreases, but does not go to zero. For high electron concentration, $R$ (blue line) is still below the Hall constant (red line) in the low electron concentration case according to Eq. (\ref{Rfree}).

\section{Conclusions}

We have numerically studied magnetotransport properties in conical helimagnets using the nonequilibrium Boltzmann equation approach. The spin component of the magnetic moment along the spiral axis is proportional to the applied magnetic field (see Eq. (\ref{sz})). This feature results in a magnetic field dependent band structure as presented in Eq. (\ref{eps}). Such field dependence in $\varepsilon(k)$ appears to be the cause of dramatic effects in spin dependent magnetoresistance and Hall constant. For spin up electrons, we have observed negative magnetoresistance, which is unexpected for single carrier types. For spin down electrons, we have found giant magnetoresistance due to the depletion of spin down electrons. For free electrons, the Hall constant described by Eq. (\ref{Rfree}) is field independent. However, for some critical value of $B$, the spin down Hall constant goes to infinity. This means that there is no Hall current, for $B > B_{critical}$. Such effect is caused by the absence of spin down carriers. For spin up carriers, we have found the Hall constant dramatically decreases with magnetic field, due to the increase in spin up charge carriers. Because of the giant spin dependent magnetoresistance and Hall resistivity, conical helimagnets could find use in spin switching devices.

\vskip 1 truein

\section*{Acknowledgment}
This work was supported by a grants from the U S National Science Foundation (No. 2228841 and No. DMR-1710512)  to the University of Wyoming. 

\bibliographystyle{apsrev}
\section*{References}
\bibliography{refs}

\begin{thebibliography}{36}
\expandafter\ifx\csname natexlab\endcsname\relax\def\natexlab#1{#1}\fi
\expandafter\ifx\csname bibnamefont\endcsname\relax
  \def\bibnamefont#1{#1}\fi
\expandafter\ifx\csname bibfnamefont\endcsname\relax
  \def\bibfnamefont#1{#1}\fi
\expandafter\ifx\csname citenamefont\endcsname\relax
  \def\citenamefont#1{#1}\fi
\expandafter\ifx\csname url\endcsname\relax
  \def\url#1{\texttt{#1}}\fi
\expandafter\ifx\csname urlprefix\endcsname\relax\def\urlprefix{URL }\fi
\providecommand{\bibinfo}[2]{#2}
\providecommand{\eprint}[2][]{\url{#2}}

\bibitem[{\citenamefont{Inoue and Maekawa}(1995)}]{inoue95}
\bibinfo{author}{\bibfnamefont{J.}~\bibnamefont{Inoue}} \bibnamefont{and} \bibinfo{author}{\bibfnamefont{S.}~\bibnamefont{Maekawa}}, \bibinfo{journal}{Phys. Rev. Lett.} \textbf{\bibinfo{volume}{74}}, \bibinfo{pages}{3407} (\bibinfo{year}{1995}).

\bibitem[{\citenamefont{Xiao et~al.}(2006)\citenamefont{Xiao, Zangwill, and Stiles}}]{xiao}
\bibinfo{author}{\bibfnamefont{J.}~\bibnamefont{Xiao}}, \bibinfo{author}{\bibfnamefont{A.}~\bibnamefont{Zangwill}}, \bibnamefont{and} \bibinfo{author}{\bibfnamefont{M.~D.} \bibnamefont{Stiles}}, \bibinfo{journal}{Phys. Rev. B} \textbf{\bibinfo{volume}{73}}, \bibinfo{pages}{054428} (\bibinfo{year}{2006}).

\bibitem[{\citenamefont{Yazyev and Katsnelson}(2008)}]{yazyev08}
\bibinfo{author}{\bibfnamefont{O.~V.} \bibnamefont{Yazyev}} \bibnamefont{and} \bibinfo{author}{\bibfnamefont{M.~I.} \bibnamefont{Katsnelson}}, \bibinfo{journal}{Phys. Rev. Lett.} \textbf{\bibinfo{volume}{100}}, \bibinfo{pages}{047209} (\bibinfo{year}{2008}).

\bibitem[{\citenamefont{Landau and Lifshitz}(1984)}]{ll}
\bibinfo{author}{\bibfnamefont{L.~D.} \bibnamefont{Landau}} \bibnamefont{and} \bibinfo{author}{\bibfnamefont{E.~M.} \bibnamefont{Lifshitz}}, \emph{\bibinfo{title}{Electrodynamics of Continuous Media}} (\bibinfo{publisher}{Pergamon}, \bibinfo{address}{New York}, \bibinfo{year}{1984}).

\bibitem[{\citenamefont{Sukhanov et~al.}(2022)\citenamefont{Sukhanov, Tymoshenko, Kulbakov, Cameron, Kocsis, Walker, Ivanov, Park, Pomjakushin, Nikitin et~al.}}]{sukh22}
\bibinfo{author}{\bibfnamefont{A.~S.} \bibnamefont{Sukhanov}}, \bibinfo{author}{\bibfnamefont{Y.~V.} \bibnamefont{Tymoshenko}}, \bibinfo{author}{\bibfnamefont{A.~A.} \bibnamefont{Kulbakov}}, \bibinfo{author}{\bibfnamefont{A.~S.} \bibnamefont{Cameron}}, \bibinfo{author}{\bibfnamefont{V.}~\bibnamefont{Kocsis}}, \bibinfo{author}{\bibfnamefont{H.~C.} \bibnamefont{Walker}}, \bibinfo{author}{\bibfnamefont{A.}~\bibnamefont{Ivanov}}, \bibinfo{author}{\bibfnamefont{J.~T.} \bibnamefont{Park}}, \bibinfo{author}{\bibfnamefont{V.}~\bibnamefont{Pomjakushin}}, \bibinfo{author}{\bibfnamefont{S.~E.} \bibnamefont{Nikitin}}, \bibnamefont{et~al.}, \bibinfo{journal}{Phys. Rev. B} \textbf{\bibinfo{volume}{105}}, \bibinfo{pages}{134424} (\bibinfo{year}{2022}).

\bibitem[{\citenamefont{Wang et~al.}(2020)\citenamefont{Wang, Su, Lin, and Batista}}]{wang20}
\bibinfo{author}{\bibfnamefont{Z.}~\bibnamefont{Wang}}, \bibinfo{author}{\bibfnamefont{Y.}~\bibnamefont{Su}}, \bibinfo{author}{\bibfnamefont{S.-Z.} \bibnamefont{Lin}}, \bibnamefont{and} \bibinfo{author}{\bibfnamefont{C.~D.} \bibnamefont{Batista}}, \bibinfo{journal}{Phys. Rev. Lett.} \textbf{\bibinfo{volume}{124}}, \bibinfo{pages}{207201} (\bibinfo{year}{2020}).

\bibitem[{\citenamefont{Ozawa et~al.}(2017)\citenamefont{Ozawa, Hayami, and Motome}}]{ozawa17}
\bibinfo{author}{\bibfnamefont{R.}~\bibnamefont{Ozawa}}, \bibinfo{author}{\bibfnamefont{S.}~\bibnamefont{Hayami}}, \bibnamefont{and} \bibinfo{author}{\bibfnamefont{Y.}~\bibnamefont{Motome}}, \bibinfo{journal}{Phys. Rev. Lett.} \textbf{\bibinfo{volume}{118}}, \bibinfo{pages}{147205} (\bibinfo{year}{2017}).

\bibitem[{\citenamefont{Hayami et~al.}(2017)\citenamefont{Hayami, Ozawa, and Motome}}]{hayami17}
\bibinfo{author}{\bibfnamefont{S.}~\bibnamefont{Hayami}}, \bibinfo{author}{\bibfnamefont{R.}~\bibnamefont{Ozawa}}, \bibnamefont{and} \bibinfo{author}{\bibfnamefont{Y.}~\bibnamefont{Motome}}, \bibinfo{journal}{Phys. Rev. B} \textbf{\bibinfo{volume}{95}}, \bibinfo{pages}{224424} (\bibinfo{year}{2017}).

\bibitem[{\citenamefont{Hayami and Motome}(2021)}]{hayami21}
\bibinfo{author}{\bibfnamefont{S.}~\bibnamefont{Hayami}} \bibnamefont{and} \bibinfo{author}{\bibfnamefont{Y.}~\bibnamefont{Motome}}, \bibinfo{journal}{Phys. Rev. B} \textbf{\bibinfo{volume}{103}}, \bibinfo{pages}{024439} (\bibinfo{year}{2021}).

\bibitem[{\citenamefont{Binz and Vishwanath}(2006)}]{binz06}
\bibinfo{author}{\bibfnamefont{B.}~\bibnamefont{Binz}} \bibnamefont{and} \bibinfo{author}{\bibfnamefont{A.}~\bibnamefont{Vishwanath}}, \bibinfo{journal}{Phys. Rev. B} \textbf{\bibinfo{volume}{74}}, \bibinfo{pages}{214408} (\bibinfo{year}{2006}).

\bibitem[{\citenamefont{Tsunoda}(1989)}]{tsunoda89}
\bibinfo{author}{\bibfnamefont{Y.}~\bibnamefont{Tsunoda}}, \bibinfo{journal}{Journal of Physics: Condensed Matter} \textbf{\bibinfo{volume}{1}}, \bibinfo{pages}{10427} (\bibinfo{year}{1989}).

\bibitem[{\citenamefont{Sandratskii and K\"ubler}(1996)}]{sandratskii96}
\bibinfo{author}{\bibfnamefont{L.}~\bibnamefont{Sandratskii}} \bibnamefont{and} \bibinfo{author}{\bibfnamefont{J.}~\bibnamefont{K\"ubler}}, \bibinfo{journal}{Physica B: Condensed Matter} \textbf{\bibinfo{volume}{217}}, \bibinfo{pages}{167} (\bibinfo{year}{1996}).

\bibitem[{\citenamefont{Kurz et~al.}(2004)\citenamefont{Kurz, F\"orster, Nordstr\"om, Bihlmayer, and Bl\"ugel}}]{kurz04}
\bibinfo{author}{\bibfnamefont{P.}~\bibnamefont{Kurz}}, \bibinfo{author}{\bibfnamefont{F.}~\bibnamefont{F\"orster}}, \bibinfo{author}{\bibfnamefont{L.}~\bibnamefont{Nordstr\"om}}, \bibinfo{author}{\bibfnamefont{G.}~\bibnamefont{Bihlmayer}}, \bibnamefont{and} \bibinfo{author}{\bibfnamefont{S.}~\bibnamefont{Bl\"ugel}}, \bibinfo{journal}{Phys. Rev. B} \textbf{\bibinfo{volume}{69}}, \bibinfo{pages}{024415} (\bibinfo{year}{2004}).

\bibitem[{\citenamefont{Wang et~al.}(2016)\citenamefont{Wang, Feng, Cheng, Wu, Luo, and Rosenbaum}}]{wang16}
\bibinfo{author}{\bibfnamefont{Y.}~\bibnamefont{Wang}}, \bibinfo{author}{\bibfnamefont{Y.}~\bibnamefont{Feng}}, \bibinfo{author}{\bibfnamefont{J.-G.} \bibnamefont{Cheng}}, \bibinfo{author}{\bibfnamefont{W.}~\bibnamefont{Wu}}, \bibinfo{author}{\bibfnamefont{J.~L.} \bibnamefont{Luo}}, \bibnamefont{and} \bibinfo{author}{\bibfnamefont{T.~F.} \bibnamefont{Rosenbaum}}, \bibinfo{journal}{Nature Communications} \textbf{\bibinfo{volume}{7}}, \bibinfo{pages}{13037} (\bibinfo{year}{2016}).

\bibitem[{\citenamefont{Nakanishi et~al.}(1980)\citenamefont{Nakanishi, Yanase, Hasegawa, and Kataoka}}]{nakanishi80}
\bibinfo{author}{\bibfnamefont{O.}~\bibnamefont{Nakanishi}}, \bibinfo{author}{\bibfnamefont{A.}~\bibnamefont{Yanase}}, \bibinfo{author}{\bibfnamefont{A.}~\bibnamefont{Hasegawa}}, \bibnamefont{and} \bibinfo{author}{\bibfnamefont{M.}~\bibnamefont{Kataoka}}, \bibinfo{journal}{Solid State Communications} \textbf{\bibinfo{volume}{35}}, \bibinfo{pages}{995} (\bibinfo{year}{1980}).

\bibitem[{\citenamefont{Maleyev}(2006)}]{maleyev06}
\bibinfo{author}{\bibfnamefont{S.~V.} \bibnamefont{Maleyev}}, \bibinfo{journal}{Phys. Rev. B} \textbf{\bibinfo{volume}{73}}, \bibinfo{pages}{174402} (\bibinfo{year}{2006}).

\bibitem[{\citenamefont{Grigoriev et~al.}(2014)\citenamefont{Grigoriev, Siegfried, Altynbayev, Potapova, Dyadkin, Moskvin, Menzel, Heinemann, Axenov, Fomicheva et~al.}}]{grigoriev14}
\bibinfo{author}{\bibfnamefont{S.~V.} \bibnamefont{Grigoriev}}, \bibinfo{author}{\bibfnamefont{S.-A.} \bibnamefont{Siegfried}}, \bibinfo{author}{\bibfnamefont{E.~V.} \bibnamefont{Altynbayev}}, \bibinfo{author}{\bibfnamefont{N.~M.} \bibnamefont{Potapova}}, \bibinfo{author}{\bibfnamefont{V.}~\bibnamefont{Dyadkin}}, \bibinfo{author}{\bibfnamefont{E.~V.} \bibnamefont{Moskvin}}, \bibinfo{author}{\bibfnamefont{D.}~\bibnamefont{Menzel}}, \bibinfo{author}{\bibfnamefont{A.}~\bibnamefont{Heinemann}}, \bibinfo{author}{\bibfnamefont{S.~N.} \bibnamefont{Axenov}}, \bibinfo{author}{\bibfnamefont{L.~N.} \bibnamefont{Fomicheva}}, \bibnamefont{et~al.}, \bibinfo{journal}{Phys. Rev. B} \textbf{\bibinfo{volume}{90}}, \bibinfo{pages}{174414} (\bibinfo{year}{2014}).

\bibitem[{\citenamefont{Nakanishi et~al.}(1983)\citenamefont{Nakanishi, Kataoka, and Yanase}}]{nakanishi83}
\bibinfo{author}{\bibfnamefont{O.}~\bibnamefont{Nakanishi}}, \bibinfo{author}{\bibfnamefont{M.}~\bibnamefont{Kataoka}}, \bibnamefont{and} \bibinfo{author}{\bibfnamefont{A.}~\bibnamefont{Yanase}}, \bibinfo{journal}{Journal of Magnetism and Magnetic Materials} \textbf{\bibinfo{volume}{31-34}}, \bibinfo{pages}{339} (\bibinfo{year}{1983}).

\bibitem[{\citenamefont{Kulikov and Tugushev}(1984)}]{kulikov84}
\bibinfo{author}{\bibfnamefont{N.~I.} \bibnamefont{Kulikov}} \bibnamefont{and} \bibinfo{author}{\bibfnamefont{V.~V.} \bibnamefont{Tugushev}}, \bibinfo{journal}{Soviet Physics Uspekhi} \textbf{\bibinfo{volume}{27}}, \bibinfo{pages}{954} (\bibinfo{year}{1984}).

\bibitem[{\citenamefont{Plumer}(1990)}]{plumer90}
\bibinfo{author}{\bibfnamefont{M.~L.} \bibnamefont{Plumer}}, \bibinfo{journal}{Journal of Physics: Condensed Matter} \textbf{\bibinfo{volume}{2}}, \bibinfo{pages}{7503} (\bibinfo{year}{1990}).

\bibitem[{\citenamefont{Pfleiderer et~al.}(2005)\citenamefont{Pfleiderer, Reznik, Pintschovius, and LĂ¶hneysen}}]{pfleiderer05}
\bibinfo{author}{\bibfnamefont{C.}~\bibnamefont{Pfleiderer}}, \bibinfo{author}{\bibfnamefont{D.}~\bibnamefont{Reznik}}, \bibinfo{author}{\bibfnamefont{L.}~\bibnamefont{Pintschovius}}, \bibnamefont{and} \bibinfo{author}{\bibfnamefont{H.}~\bibnamefont{LĂ¶hneysen}}, \bibinfo{journal}{Physica B: Condensed Matter} \textbf{\bibinfo{volume}{359-361}}, \bibinfo{pages}{1159} (\bibinfo{year}{2005}).

\bibitem[{\citenamefont{Schmidt et~al.}(2016)\citenamefont{Schmidt, Hagemeister, Hsu, Kubetzka, von Bergmann, and Wiesendanger}}]{schmidt16}
\bibinfo{author}{\bibfnamefont{L.}~\bibnamefont{Schmidt}}, \bibinfo{author}{\bibfnamefont{J.}~\bibnamefont{Hagemeister}}, \bibinfo{author}{\bibfnamefont{P.-J.} \bibnamefont{Hsu}}, \bibinfo{author}{\bibfnamefont{A.}~\bibnamefont{Kubetzka}}, \bibinfo{author}{\bibfnamefont{K.}~\bibnamefont{von Bergmann}}, \bibnamefont{and} \bibinfo{author}{\bibfnamefont{R.}~\bibnamefont{Wiesendanger}}, \bibinfo{journal}{New Journal of Physics} \textbf{\bibinfo{volume}{18}}, \bibinfo{pages}{075007} (\bibinfo{year}{2016}).

\bibitem[{\citenamefont{Yang et~al.}(2021)\citenamefont{Yang, Naaman, Paltiel, and Parkin}}]{Yang21}
\bibinfo{author}{\bibfnamefont{S.-H.} \bibnamefont{Yang}}, \bibinfo{author}{\bibfnamefont{R.}~\bibnamefont{Naaman}}, \bibinfo{author}{\bibfnamefont{Y.}~\bibnamefont{Paltiel}}, \bibnamefont{and} \bibinfo{author}{\bibfnamefont{S.~S.~P.} \bibnamefont{Parkin}}, \bibinfo{journal}{Nature Reviews Physics} \textbf{\bibinfo{volume}{3}}, \bibinfo{pages}{328 } (\bibinfo{year}{2021}).

\bibitem[{\citenamefont{Jiang et~al.}(2020)\citenamefont{Jiang, Nii, Arisawa, Saitoh, and Onose}}]{jiang2020}
\bibinfo{author}{\bibfnamefont{N.}~\bibnamefont{Jiang}}, \bibinfo{author}{\bibfnamefont{Y.}~\bibnamefont{Nii}}, \bibinfo{author}{\bibfnamefont{H.}~\bibnamefont{Arisawa}}, \bibinfo{author}{\bibfnamefont{E.}~\bibnamefont{Saitoh}}, \bibnamefont{and} \bibinfo{author}{\bibfnamefont{Y.}~\bibnamefont{Onose}}, \bibinfo{journal}{Nature Communications} \textbf{\bibinfo{volume}{11}}, \bibinfo{pages}{1601} (\bibinfo{year}{2020}).

\bibitem[{\citenamefont{Mohanta et~al.}(2020)\citenamefont{Mohanta, Okamoto, and Dagotto}}]{mohanta20}
\bibinfo{author}{\bibfnamefont{N.}~\bibnamefont{Mohanta}}, \bibinfo{author}{\bibfnamefont{S.}~\bibnamefont{Okamoto}}, \bibnamefont{and} \bibinfo{author}{\bibfnamefont{E.}~\bibnamefont{Dagotto}}, \bibinfo{journal}{Phys. Rev. B} \textbf{\bibinfo{volume}{102}}, \bibinfo{pages}{064430} (\bibinfo{year}{2020}).

\bibitem[{\citenamefont{Ustinov and Yasyulevich}(2020)}]{ust}
\bibinfo{author}{\bibfnamefont{V.~V.} \bibnamefont{Ustinov}} \bibnamefont{and} \bibinfo{author}{\bibfnamefont{I.~A.} \bibnamefont{Yasyulevich}}, \bibinfo{journal}{Phys. Rev. B} \textbf{\bibinfo{volume}{102}}, \bibinfo{pages}{134431} (\bibinfo{year}{2020}).

\bibitem[{\citenamefont{Goto et~al.}(2021)\citenamefont{Goto, Ishihara, and Yokoshi}}]{goto21}
\bibinfo{author}{\bibfnamefont{Y.}~\bibnamefont{Goto}}, \bibinfo{author}{\bibfnamefont{H.}~\bibnamefont{Ishihara}}, \bibnamefont{and} \bibinfo{author}{\bibfnamefont{N.}~\bibnamefont{Yokoshi}}, \bibinfo{journal}{Japanese Journal of Applied Physics} \textbf{\bibinfo{volume}{60}}, \bibinfo{pages}{098001} (\bibinfo{year}{2021}).

\bibitem[{\citenamefont{Zadorozhnyi and Dahnovsky}(2023)}]{ZD4}
\bibinfo{author}{\bibfnamefont{A.}~\bibnamefont{Zadorozhnyi}} \bibnamefont{and} \bibinfo{author}{\bibfnamefont{Y.}~\bibnamefont{Dahnovsky}}, \bibinfo{journal}{Phys. Rev. B} \textbf{\bibinfo{volume}{107}}, \bibinfo{pages}{035202} (\bibinfo{year}{2023}).

\bibitem[{\citenamefont{{Zadorozhnyi} et~al.}(2023)\citenamefont{{Zadorozhnyi}, {Rivlis}, and {Dahnovsky}}}]{Zadorozhnyi_2023}
\bibinfo{author}{\bibfnamefont{A.}~\bibnamefont{{Zadorozhnyi}}}, \bibinfo{author}{\bibfnamefont{R.}~\bibnamefont{{Rivlis}}}, \bibnamefont{and} \bibinfo{author}{\bibfnamefont{Y.}~\bibnamefont{{Dahnovsky}}}, \bibinfo{journal}{Physics Review B} \textbf{\bibinfo{volume}{108}}, \bibinfo{eid}{014405} (\bibinfo{year}{2023}).

\bibitem[{\citenamefont{Sandratskii}(1986)}]{sandratskii86}
\bibinfo{author}{\bibfnamefont{L.~M.} \bibnamefont{Sandratskii}}, \bibinfo{journal}{physica status solidi (b)} \textbf{\bibinfo{volume}{136}}, \bibinfo{pages}{167} (\bibinfo{year}{1986}).

\bibitem[{\citenamefont{Kurebayashi and Nagaosa}(2021)}]{kurebayashi21}
\bibinfo{author}{\bibfnamefont{D.}~\bibnamefont{Kurebayashi}} \bibnamefont{and} \bibinfo{author}{\bibfnamefont{N.}~\bibnamefont{Nagaosa}}, \bibinfo{journal}{Communications Physics} \textbf{\bibinfo{volume}{4}}, \bibinfo{pages}{260} (\bibinfo{year}{2021}).

\bibitem[{\citenamefont{Bruno et~al.}(2004)\citenamefont{Bruno, Dugaev, and Taillefumier}}]{bruno04}
\bibinfo{author}{\bibfnamefont{P.}~\bibnamefont{Bruno}}, \bibinfo{author}{\bibfnamefont{V.~K.} \bibnamefont{Dugaev}}, \bibnamefont{and} \bibinfo{author}{\bibfnamefont{M.}~\bibnamefont{Taillefumier}}, \bibinfo{journal}{Phys. Rev. Lett.} \textbf{\bibinfo{volume}{93}}, \bibinfo{pages}{096806} (\bibinfo{year}{2004}).

\bibitem[{\citenamefont{Everschor-Sitte and Sitte}(2014)}]{sitte14}
\bibinfo{author}{\bibfnamefont{K.}~\bibnamefont{Everschor-Sitte}} \bibnamefont{and} \bibinfo{author}{\bibfnamefont{M.}~\bibnamefont{Sitte}}, \bibinfo{journal}{Journal of Applied Physics} \textbf{\bibinfo{volume}{115}}, \bibinfo{pages}{172602} (\bibinfo{year}{2014}).

\bibitem[{\citenamefont{Anselm}(1981)}]{anselm}
\bibinfo{author}{\bibfnamefont{A.}~\bibnamefont{Anselm}}, \emph{\bibinfo{title}{Introduction to Semiconductor Theory}} (\bibinfo{publisher}{Mir, Moscow}, \bibinfo{year}{1981}).

\bibitem[{\citenamefont{Nereson et~al.}(1964)\citenamefont{Nereson, Olsen, and Arnold}}]{olsen64}
\bibinfo{author}{\bibfnamefont{N.~G.} \bibnamefont{Nereson}}, \bibinfo{author}{\bibfnamefont{C.~E.} \bibnamefont{Olsen}}, \bibnamefont{and} \bibinfo{author}{\bibfnamefont{G.~P.} \bibnamefont{Arnold}}, \bibinfo{journal}{Phys. Rev.} \textbf{\bibinfo{volume}{135}}, \bibinfo{pages}{A176} (\bibinfo{year}{1964}).

\bibitem[{\citenamefont{Askerov}(1970)}]{askerov70}
\bibinfo{author}{\bibfnamefont{B.}~\bibnamefont{Askerov}}, \emph{\bibinfo{title}{Kinetic effects in semiconductors (in Russian)}} (\bibinfo{publisher}{Nauka, Leningrad}, \bibinfo{year}{1970}).

\end{thebibliography}

\end{document}